\documentclass{llncs}
\pdfoutput=1
\usepackage{etex}
\usepackage{algorithm,algorithmic}
\usepackage{xargs,ifthen}

\usepackage[utf8]{inputenc}
\usepackage[T1]{fontenc}
\usepackage[english]{babel} 

\usepackage{enumitem}
        \setitemize{itemsep=1pt,topsep=3pt,parsep=0pt,partopsep=0pt}
\usepackage{hyperref} 
\usepackage{xcolor}
\usepackage{framed} 
\usepackage{blindtext} 
\usepackage{multicol} 
\usepackage{wrapfig} 

\usepackage{amsmath,amssymb}
\usepackage{microtype}

\usepackage[all]{xy} 
\usepackage{stmaryrd}
\usepackage{cmll}

\usepackage{graphicx}

\hypersetup{
	plainpages			=	false,		
	bookmarksnumbered	=	true,		
	bookmarksopen,
	bookmarksdepth=2	,
	colorlinks			=	true, 		
	breaklinks			=	true,		
	linkcolor			=	red,	 	
	pdfdisplaydoctitle	= 	true,		
}

\usepackage{mathtools}
\usepackage{csquotes}

\newtheorem{theorm}{Theorem}
\newtheorem{propositin}[theorm]{Proposition}
\newtheorem{definitin}[theorm]{Definition}
\newtheorem{lema}[theorm]{Lemma}
\newtheorem{corollay}[theorm]{Corollary}

\newcommand{\optname}[1]{\ifthenelse{\equal{#1}{}}{\!\!\textbf\\}{\!\!\textbf{(#1)\\}}}
\newcommand{\optparen}[1]{\ifthenelse{\equal{#1}{}}{}{~(#1)}}

\newcommandx{\define}[2][1=]{
			\setlength{\parindent}{0pt}
	\begin{definitin}\optname{#1}
			\vspace{0.7pt}
			#2
			\vspace{0.7pt} 
	\end{definitin}
}

\renewcommandx{\proposition}[2][1=]{
			\setlength{\parindent}{0pt}
	\begin{propositin}\optname{#1}
			\vspace{0.7pt}
			#2
			\vspace{0.7pt} 
	\end{propositin}
}

\renewcommandx{\theorem}[2][1=]{
			\setlength{\parindent}{0pt}
	\begin{theorm}\optname{#1}
			\vspace{0.7pt}
			#2
			\vspace{0.7pt} 
	\end{theorm}
}

\renewcommandx{\lemma}[2][1=]{
			\setlength{\parindent}{0pt}
	\begin{lema}\optname{#1}
			\vspace{0.7pt}
			\noindent
			#2
			\vspace{0.7pt} 
	\end{lema}
}

\renewcommandx{\corollary}[2][1=]{
			\setlength{\parindent}{0pt}
	\begin{corollay}\optname{#1}
			\vspace{0.7pt}
			\noindent
			#2
			\vspace{0.7pt} 
	\end{corollay}
}

\renewcommandx{\proof}[2][1=]{\noindent\textit{Proof\optparen{#1}}. #2\qed}

\newcommand{\tpower}[1]{^{\otimes #1}}
\newcommand{\ptensor}{\,\dot{\otimes}\,}
\newcommand{\unitalg}{\C I}
\newcommand{\unit}{I}

\newcommand{\comp}{\TT{Comp}}

\renewcommand{\i}{\TT{R}}
\renewcommand{\o}{\TT{L}}
\newcommand{\set}[2]{\left\{\:#1 \ \middle|\ #2\:\right\}}
\newcommand{\vars}{\TT V}

\newcommand{\terms}{\TT T}
\newcommand{\lang}{\mathcal{L}}
\newcommand{\closed}{\TT T_{\TT c}}
\newcommand{\closedh}{\BB V_{\TT c}}
\newcommand{\ualg}{\C U}

\newcommand{\vect}{\TT{Vect}}

\newcommand{\start}{\star}
\newcommand{\IO}{\TT{LR}}
\newcommand{\p}{\scalebox{0.6}{$\,\bullet\,$}}

\newcommand{\concrete}{+}
\newcommand{\C}[1]{\mathcal{#1}}
\newcommand{\F}[1]{\mathfrak{#1}}
\newcommand{\BB}[1]{\mathbb{#1}}

\newcommand{\TT}[1]{\mathtt{#1}}

\newcommand{\Id}{\mbox{Id}}

\newcommand{\flow}{\,\scalebox{1.1}{$\leftharpoonup$}\,}
\newcommand{\flows}{\C F}

\newcommand{\sflow}{\,\scalebox{1.1}{$\leftrightharpoons$}\,}

\newcommand{\trans}[2]{\xrightarrow[\,#2\,]{\,#1\,}}
\newcommand{\dom}{\TT{Dom}}
\newcommand{\var}{\TT{Var}}

\newcommand{\notations}[1]{\noindent\textbf{Notations.} {\it#1}}
\newcommand{\notation}[1]{\noindent\textbf{Notation.} {\it#1}}
\renewcommand{\remark}[1]{\noindent\textbf{Remark.} {\it #1}}

\renewcommand{\example}[1]{\noindent{\textbf{Example.} \it#1}}
\newcommand{\examples}[1]{\noindent{\textbf{Examples.} \it #1}}

\begin{document}
	\setlength{\parindent}{13pt} 
	\renewcommand{\labelitemi}{\scalebox{0.8}{$\bullet$}} 
	
	\hyphenpenalty=2000
	\tolerance=600

\title{Unification and Logarithmic Space}
\author{Clément Aubert \and Marc Bagnol\thanks{This work was partly supported by the ANR-10-BLAN-0213 Logoi and the ANR-11-BS02-0010 Récré.}}
\institute{Aix-Marseille Université, CNRS, I2M, UMR 7373, 13453 Marseille, France}

\maketitle




\begin{abstract}
We present an algebraic characterization of the complexity classes \textsc{Logspace} and \textsc{NLogspace}, using an algebra with a composition law based on unification. This new bridge between unification and complexity classes is inspired from proof theory and more specifically linear logic and Geometry of Interaction.

We show how unification can be used to build a model of computation by means of specific subalgebras associated to finite permutations groups.

We then prove that whether an observation (the algebraic counterpart of a program) accepts a word can be decided within logarithmic space.
We also show that the construction can naturally represent pointer machines, an intuitive way of understanding logarithmic space computing.


\end{abstract}


\section*{Introduction}
\addcontentsline{toc}{section}{Introduction}
\textbf{Proof theory and complexity theory.}
There is a longstanding tradition of relating proof theory (more specifically linear logic \cite{girard_linear_1987}) and implicit complexity theory that dates back to the introduction of bounded \cite{Girard1992} and light~\cite{girard_light_1994} logics.
Control over the modalities \cite{schopp_stratified_2007,Lago2010}, 
type assignment \cite{gaboardi_logical_2008} and stratification of exponential boxes \cite{baillot_linear_2010}, to name a few, led to a clearer understanding of the complexity bounds linear logic could entail on the cut-elimination procedure.

We propose to push further this approach by adopting a more semantical and algebraic point of view that will allow us to capture non-deterministic logarithmic space computation.

\smallskip\noindent
\textbf{Geometry of Interaction.}
As the study of cut-elimination has grown as a central topic in proof theory, its mathematical modeling became of great interest.
The Geometry of Interaction \cite{girard_towards_1989} research program led to mathematical models of cut-elimination in terms of paths in proofnets~\cite{asperti_paths_1994}, token machines \cite{laurent_token_2001} and operator algebras \cite{girard_geometry_1989}.
It was already used with complexity concerns \cite{baillot_elementary_2001,girard_normativity_2012}. 

Recent works \cite{girard_normativity_2012,aubert_characterizing_2012,seiller_logarithmic_2013} studied the link between Geometry of Interaction and logarithmic space, relying on the theory of von Neumann algebras.
Those three articles are indubitably sources of inspiration of this work, but the whole construction is made anew, in a simpler framework.
%

\smallskip\noindent
\textbf{Unification.} 
Unification is one of the key-concepts of theoretical computer science, for it is used in logic programming and is a classical subject of study for complexity theory.
It was shown \cite{girard_geometry_1995,girard_three_lightings} that one can model \hbox{cut-elimination} with unification techniques.
%

Execution will be expressed in terms of matching in a \emph{unification algebra}.
This is a simple framework, yet expressive enough to encode the action of finite permutation groups on an unbounded tensor product, which is a crucial ingredient of our construction.
%



\smallskip

\bigskip\noindent
\textbf{Contribution.}
We carry on the methodology of bridging Geometry of Interaction and complexity theory with a renewed approach.
It relies on an simpler representation of execution in a unification-based algebra, proved to capture exactly logarithmic space complexity.

While the representation of inputs (words over a finite alphabet) comes from the classical Church representation of lists, observations (the algebraic counterpart of programs) are shown to correspond  very naturally to a notion of pointer machines.
This correspondence allows us to prove that reversibility (of machines) is related to the algebraic notion of isometricity (of observations).




\smallskip\noindent
\textbf{Organization of this article.} In Sect.\ref{sec_unification} we review some classical results on unification of first-order terms and use them to build the algebra that will constitute our computational setting.

We explain in Sect.\ref{sec_words} how words and computing devices (observations) can be modeled by particular elements of this algebra.
The way they interact to yield a notion of language recognized by an observation is described in Sect.\ref{sec_normativity}.

Finally, we show in Sect.\ref{sec_logspace} that our construction captures exactly logarithmic space computation, both deterministic and non-deterministic.

\vspace{-2mm}

\section{The Unification Algebra}\label{sec_unification}
	\vspace{-1mm}
	\subsection{Unification}
	Unification can be generally thought of as the study of formal solving of equations between terms.

This topic was introduced by Herbrand, but became really widespread after the work of J.~A.~Robinson on automated theorem proving.
The unification technique is also at the core of the logic programming language \textsc{Prolog} and type inference for functional programming languages such as \textsc{CaML} and \textsc{Haskell}.


Specifically, we will be interested in the following problem: 
\vspace{-1mm}
\begin{center}
	\it Given two terms, can they be \enquote{made equal} by replacing their variables?
\end{center}
\vspace{-2mm}
\define[terms]
	{We consider the following set of first-order \emph{terms}
	\[\terms::=\ \ x,y,z,\:\dots\ |\ \TT a,\tt b,\tt c,\:\dots\ |\ \terms\p\terms\]
	where  $\,x,y,z,\:\dots\in \vars\,$ are variables, $\,\TT a,\TT b,\TT c,\:\dots\,$ are constants and $\,\p\,$ is a binary function symbol.

\smallskip
For any $\,t\in\terms\,$, we will write $\,\var(t)\,$ the set of variables occurring in $\,t\,$.
We say that a term is closed when $\,\var(t)=\varnothing\,$, and denote $\,\closed\,$ the set of closed terms. 
}

\notation{The binary function symbol $\,\p\,$ is not associative, but we will write it by convention as \emph{right associating} to lighten notations: $\:t\p u\p v \::=\:t\p(u\p v) \,$}


\define[substitution]
	{A \emph{substitution} is a map $\,\theta:\:\vars \rightarrow \terms\,$ such that the set $\,\dom(\theta):=\{\:v\in \vars\:|\:\theta(v)\not=v\:\}\,$ (the \emph{domain} of $\,\theta\,$) is finite.
	A substitution with domain $\,\{\,x_1,\dots,x_n\,\}\,$ such that $\,\theta(x_1)=u_1\,,\,\dots\,,\,\theta(x_n)=u_n\,$ will be written as $\,\{\:x_1\mapsto u_1\,;\,\dots\,;\, x_n\mapsto u_n\:\}\,$.

\smallskip
	If $\,t\in \terms\,$ is a term we write $\,t.\theta\,$ the term $\,t\,$ where any occurrence of any variable $\,x\,$ has been replaced by $\,\theta(x)\,$.
	
	\smallskip
	If $\,\theta=\{\:x_i\mapsto u_i\:\}$ and $\,\psi=\{\:y_j\mapsto v_j\:\}\,$, their \emph{composition} is defined as
\[\theta;\psi\::=\: 
	\{\:x_i\mapsto u_i.\psi\:\} \:\cup\:
	\{\:y_j\mapsto v_j \:|\;y_j \not\in \dom(\theta)\:\}\]
\vspace{-7mm}
}

\remark{The composition of substitutions is such that $\,t.(\theta;\psi)=(t.\theta).\psi\,$ holds.}

\define[renamings and instances]
{\label{renaming}A \emph{renaming} is a substitution $\,\alpha\,$ such that $\,\alpha(\vars)\subseteq \vars\,$ and that is bijective. A term $\,t'\,$ is a \emph{renaming} of $\,t\,$ if $\,t'=t.\alpha\,$ for some renaming $\,\alpha\,$.

\smallskip Two substitutions $\,\theta,\psi\,$ are equal \emph{up to renaming} if there is a renaming $\,\alpha\,$ such that $\,\psi=\theta;\alpha\,$.

\smallskip A substitution $\,\psi\,$ is an \emph{instance} of $\,\theta\,$ if there is a substitution $\,\sigma\,$ such that $\,\psi=\theta;\sigma\,$.

}

\proposition
	{\label{folklore}Let $\,\theta,\psi\,$ be two substitutions. If $\,\theta\,$ is an instance of $\,\psi\,$ and $\,\psi\,$ is an instance of $\,\theta\,$, then they are equal up to renaming.
}



\define[unification]
	{\label{def_unification}Two terms $\,t,u\,$ are \emph{unifiable} if there is a substitution $\,\theta\,$ such that $\,t.\theta=u.\theta\,$.
	
	\smallskip	
	We say that $\theta\,$ is a \emph{most general unifier (MGU) of $\,t,u\,$} if any other unifier of $\,t,u\,$ is an instance of $\,\theta\,$.
}

\remark{It follows from Proposition \ref{folklore} that any two MGU of a pair of terms are equal up to renaming.}

\smallskip
We will be interested mostly in the weaker variant of unification where one can first perform renamings on terms so that their variables are distinct, we introduce therefore a specific vocabulary for it.

\define[disjointness and matching]
	{\label{disjoint}Two terms $\,t,u\,$ are \emph{matchable} if $\,t',u'\,$ are unifiable, where $\,t',u'\,$ are renamings (Definition \ref{renaming}) of $\,t,u\,$ such that $\,\var(t')\cap\var(u')=\varnothing\,$.
	
	If two terms are not matchable, they are said to be \emph{disjoint}.
}

\example{$x\,$ and $\,\TT f\p x\,$ are not unifiable.

But they are matchable, as $\,x.\{\,x\mapsto y\,;\,y\mapsto x\,\}=y\,$ which is unifiable with $\,\TT f\p x\,$.

\smallskip
More generally, disjointness is stronger than non-unifiability.}

\medskip
The crucial feature of first-order unification is the (decidable) existence of most general unifiers for unification problems that have a solution.

\proposition[MGU]
{If a unification problem has a unifier, then it has a MGU.

Whether two terms are unifiable and, in case they are, finding a MGU is decidable.}

As unification grew in importance, the study of its complexity gained in attention.
A complete survey \cite{knight_unification_1989} tells the story of the bounds getting sharpened: general first-order unification was finally proved \cite{dwork_sequential_1984} to be a \textsc{Ptime}-complete problem.

\smallskip
In this article, we are concerned with a very much simpler case of the problem: the matching (Definition \ref{disjoint}) of linear terms (\textit{ie.} where variables occur at most once).
This case can be solved in a space-efficient way.

\proposition[matching in logarithmic space {\cite[Lemma 20]{dwork_parallel_1988}}]
{\label{unif-logspace}%
Whether two linear terms $\,t,u\,$ with disjoint sets of variables are unifiable, and if so finding a MGU, can be computed in logarithmic space in the size\footnote{The \emph{size} of a term is the total number of occurrences of symbols in it.} of $\,t, u\,$ on a deterministic Turing machine 
}
\vspace{-2mm}
The lemma in \cite{dwork_parallel_1988} actually states that the problem is in \textsc{NC$^1$}, a complexity class of parallel computation known to be included in \textsc{Logspace}.

We will use only a special case of the result, matching a linear term against a closed term.
\vspace{-2mm}

	\subsection{Flows and Wirings}
	We now use the notions we just saw to build an algebra with a product based on unification.
Let us start with a monoid with a partially defined product, which will be the basis of the construction.

\define[flows]
	{\label{def_flow}A \emph{flow} is an oriented pair written $\,t\flow u\,$ with $\,t,u\in\terms\,$ such that $\,\var(t)=\var(u)\,$.
	
	Flows are considered up to renaming: for any renaming $\,\alpha\,$,
	$\,t\flow u\,=\,t.\alpha\flow u.\alpha\,$.

	\smallskip
	We will write $\,\flows\,$ the set of (equivalence classes of) flows.
	
	We set $\,\unit:=x\flow x\,$ and $\,(t\flow u)^\dagger:=u\flow t\,$ so that $\,(.)^\dagger\,$ is an involution of $\,\flows\,$.
	}

A flow $\,t\flow u\,$ can be thought of as a \texttt{ `match ... with u -> t' } in a ML-style language.
The composition of flows follows this intuition.
\define[product of flows]
	{Let $\,u\flow v\in \flows\,$ and $\,t\flow w\in \flows\,$.
	Suppose we have chosen two representatives of the renaming classes such that their sets of variables are disjoint.

The \emph{product} of $\,u\flow v\,$ and $\,t\flow w\,$ is defined
if $\,v,t\,$ are unifiable with MGU $\,\theta\,$ (the choice of a MGU does not matter because of the remark following Definition~\ref{def_unification}) and in that case:
$\,(u\flow v)(t\flow w)\,:=\:u.\theta \flow w.\theta\,$.
}
\vspace{-3mm}
\define[action on closed terms]
	{\label{flow-action}If $\,t\in \closed\,$ is a closed term, $\,(u\flow v)(t)\,$ is defined whenever $\,t\,$ and $\,v\,$ are unifiable, with MGU $\,\theta\,$, in that case $\,(u\flow v)(t):=u.\theta\,$
}

\vspace{-1mm}
\examples{Composition of flows: $\,(x\p \TT c \flow (\TT c\p\TT c) \p x)(y\p z\flow z\p y)= x\p\TT  c \flow x\p\TT c\p\TT c\,$.

Action on a closed term: $(x\p\TT  c \flow x\p\TT c\p\TT c)(\TT d\p\TT c\p\TT c)=\TT d\p\TT c\,$.}

\remark{The condition on variables ensures that the result is a closed term (because \hbox{$\,\var(u)\subseteq\var(v)\,$}) and that the action is injective on its domain of definition (because \hbox{$\,\var(v)\subseteq\var(u)\,$}).
Moreover, the action is compatible with the product of flows: $\,l(k(t))=(l\,k)(t)\,$ and both are defined at the same time.
}

\medskip
By adding a formal element $\,\bot\,$ (representing the failure of unification) to the set of flows, one could turn the product into a completely defined operation, making $\,\C F\,$ an \emph{inverse monoid}. However, we will need to consider the wider algebra of \emph{sums} of flows that is easily defined directly from the partially defined product.

\define[wirings]
	{\emph{Wirings} are $\BB C$-linear combinations of elements of $\,\flows\,$ (formally: almost-everywhere null functions from $\,\flows\,$ to $\,\BB C\,$), endowed with the following operations:
	\vspace{-1mm}
	\[\bigg(\sum_i \lambda_i\,l_i\bigg) \bigg(\sum_j \mu_j\,k_j\bigg):=
		\:\sum_{\mathclap{\substack{i,j \,\text{ such that} \\ (l_ik_j)\,\text{is defined}}}}\lambda_i\mu_j(l_i\,k_j)
		\qquad\scriptstyle\text{(with }\,\lambda_i,\mu_j\in\BB C\,\text{ and }\,l_i,k_j\in \flows\,\text{)}\]
	\vspace{-4mm}
	\[\text{and\qquad} \bigg(\sum_i \lambda_i\,l_i\bigg)^\dagger:=\:\sum_i \,\overline\lambda_i\,l_i^\dagger
	\qquad\scriptstyle\text{(where }\,\overline\lambda\,\text{ is the complex conjugate of }\,\lambda\,\text{)}
	\]
	
	\vspace{-1mm}
	We write $\,\ualg\,$ the set of wirings and refer to it as the \emph{unification algebra}.
}

\vspace{-1mm}
\remark{Indeed, $\ualg\,$ is a unital $\ast$-algebra: it is a $\BB C$-algebra (considering the product defined above) with an involution $\,(.)^\dagger\,$ and a unit $\,\unit\,$.}

\define[partial isometries]
	{\label{piso}A \emph{partial isometry} is a wiring $\,U\in \ualg\,$ satisfying $\,UU^\dagger U=U\,$.
}

\vspace{-2mm}
\example{$\,(\TT c\p x \flow x\p\TT d)+(\TT d\p\TT c\flow \TT c\p \TT c)\,$ is a partial isometry.}

\smallskip
While $\,\ualg\,$ offers the general algebraic background to work in, we will need to consider particular kind of wirings to study computation.

\define[concrete and isometric wirings]
{\label{concrete}A wiring is \emph{concrete} whenever it is a sum of flows with all coefficients equal to~$\,1\,$.

An \emph{isometric wiring} is a concrete wiring that is also a partial isometry.

\smallskip
Given a set of wirings $\,E\,$ we write $\,E^+\,$ for the set of all concrete wirings of $\,E\,$.
}
\vspace{-1mm}
Isometric wirings enjoy a direct characterization.

\proposition[isometric wirings]
{The isometric wirings are exactly the wirings of the form
	$\,\sum_i \, u_i \flow t_i\,$
	with the $\,u_i\,$ pairwise disjoint (Definition \ref{disjoint}) and $\,t_i\,$ pairwise disjoint.
}
\vspace{-2mm}

It will be useful to consider the action of wirings on closed terms.
For this purpose we extend Definition \ref{flow-action} to wirings.

\define[action on closed terms]
	{\label{act_vect}
	Let $\,\closedh\,$ be the free $\BB C$-vector space over $\,\closed\,$.
	
	Wirings act on base vectors of $\,\closedh\,$ in the following way
	\vspace{-1mm}
	$$\bigg(\sum_i \lambda_i\,l_i\bigg)(t) :=\!\sum_{\mathclap{\substack{i \,\text{ such that} \\ l_i(t)\,\text{ is defined}}}}\lambda_i \big(l_i(t)\big) \ \ \in\: \closedh$$

\vspace{-2mm} 
which extends by linearity into an action on the whole $\,\closedh\,$.
}
\vspace{-2mm}
Isometric wirings have a particular behavior in terms of this action.

\lemma[isometric action]
{\label{lem_isom}Let $\,F\,$ be an isometric wiring and $\,t\,$ a closed term.
We have that $\,F(t)\,$ and $\,F^\dagger(t)\,$ are either $\,0\,$ or another closed term $\,t'\,$ (seen as an element of $\,\closedh\,$).}

\vspace{-5mm}

	\subsection{Tensor Product and Permutations}\label{permutation}
	We define now the representation in $\,\ualg\,$ of structures that provide enough expressivity to model computation.

Unbounded tensor products will allow to represent data of arbitrary size, and finite-support permutations will be used to manipulate these data.

\medskip
\notations{Given any set of wirings or closed terms $\,E\,$, we write $\,\vect(E)\,$ the vector space
generated by $\,E\,$, \emph{ie.} the set of \emph{finite} linear combinations of elements of $\,E\,$ (for instance $\,\vect(\closed)=\closedh\,$).

Moreover, we set 
$\,\unitalg:=\set{\lambda\unit}{\lambda\in \BB C\,}\,$ (with $\,\unit=x\flow x\,$ as in Definition~\ref{def_flow}) which is the $\ast$-algebra of multiples of the identity,
and
$\,u\sflow v:=\, u\flow v +v\flow u\,$.

\smallskip
For brevity we write \enquote{$\ast$-algebra} instead of the more correct \enquote{$\ast$-subalgebra of $\,\ualg\,$}
(\emph{ie.} a subset of $\,\ualg\,$ that is stable by linear combinations, product and $\,(.)^\dagger\,$).

}


\define[tensor product]
	{\label{ptensor}
	Let $\,u\flow v\,$ and $\,t\flow w\,$ be two flows.
	Suppose we have chosen representatives of these renaming classes that have their sets of variables disjoint. We define their \emph{tensor product} as
	$\,(u\flow v) \ptensor (t\flow w):=\: u\p t \flow v\p w\,$.
	The operation is extended to wirings by bilinearity.
	
	\smallskip
	Given two $\ast$-algebras $\,\C A,\C B\,$, we define their tensor product as the $\ast$-algebra
	\[\C A\ptensor \C B:= \,\vect\set{F\ptensor G}{F\in\C A,\:G\in \C B}\]
	\vspace{-7mm}
}

This actually defines an embedding of the algebraic tensor product of $\ast$-algebras into $\,\ualg\,$, which means in particular that $\,(F\ptensor G)(P\ptensor Q)=(FP)\ptensor(GQ)\,$. It ensures also that the $\,\ptensor\,$ operation indeed yields $\ast$-algebras.

\smallskip
\notation{As $\,\p\,$, the $\,\ptensor\,$ operation is not associative. We carry on our convention and write it as \emph{right associating}: $\,\C A\ptensor\C B\ptensor\C C \::=\:\C A\ptensor(\C B\ptensor\C C) \,$.}
\vspace{-1mm}

\define[unbounded tensor]
	{\label{unbounded}%
	Let $\,\C A\,$ be a $\ast$-algebra.
	We define the $\ast$-algebras $\,\C A\tpower n\,$ for all $\,n\in \BB N\,$ as
	\vspace{-2mm}
	\[\C A\tpower 0:= \unitalg \quad\text{and}\quad \C A\tpower {n+1}:=\,\C A \ptensor \C A\tpower n\]
	\vspace{-3mm}
	
	and the $\ast$-algebra $\ \displaystyle\C A\tpower \infty:=\:\bigcup_{\mathclap{n\in \BB N}} \:\C A\tpower n\ $.
}

We will consider finite permutations, but allow them to be composed even when their domain of definition do not match.

\smallskip
\notations{Let $\,\F S_n\,$ be the set of finite permutations over $\{1, \hdots, n\}$, if $\,\sigma\in \F S_n\,$, we define $\,\sigma_{+k}\in \F S_{n+k}\,$ as the permutation $\,\sigma\,$ extended to $\,\{\:1,\dots,n,\dots,n+k\:\}\,$ letting $\,\sigma_{+k}(n+i):=n+i\,$ for $\,i\in\{\,1,\dots,k\,\}\,$.

We also write $\,\unit_k:=\Id_{\{1,\dots,k\}}\in \F S_k\,$.}

\define[representation]
	{To a permutation $\,\sigma \in \F S_n\,$ we associate the flow
	\vspace{-1mm}
	\[[\sigma]:= \,x_1\p x_2\p\,\cdots\,\p x_n \p y\flow x_{\sigma(1)}\p x_{\sigma(2)}\p\,\cdots\,\p x_{\sigma(n)}\p y\]
	\vspace{-6mm}
	
}

A permutation $\,\sigma\in\F S_n\,$ will act on the first $\,n\,$ components of the unbounded tensor product (Definition \ref{unbounded}) by swapping them and leaving the rest unchanged.

The wirings $\,[\sigma]\,$ internalize this action: in the above definition, the variable $\,y\,$ at the end stands for the components that are not affected.

\smallskip
\example{Let $\,\tau\in\F S_2\,$ be the  permutation swapping the two elements of $\,\{1,2\}\,$ and $\,U_1\ptensor U_2\ptensor U_3\ptensor \unit \in \ualg\tpower 3\subseteq \ualg\tpower\infty\,$. We have $\,[\tau]=\,x_1\p x_2\p y\flow x_2\p x_1\p y\,$ and $\,[\tau](U_1\ptensor U_2\ptensor U_3\ptensor \unit)[\tau]^\dagger=U_2\ptensor U_1\ptensor U_3\ptensor \unit\,$.}

\proposition[representation]
	{\label{open-represent}For $\,\sigma\in \F S_n\,$ and $\,\tau\in\F S_{n+k}\,$ we have
	\vspace{-1mm}
	\[[\sigma_{+k}]=[\sigma][\unit_{n+k}]=[\unit_{n+k}][\sigma] \qquad
	[\sigma_{+k}\circ\tau]=[\sigma][\tau]\qquad
	 \text{and} \qquad
	 [\sigma^{-1}]=[\sigma]^\dagger
	\]
	\vspace{-5mm}
}

\define[permutation algebra]
	{For $\,n\in \BB N\,$ we set $\:[\F S_n]:=\,\set{[\sigma]}{\sigma \in \F S_n}\,$ and $\,\C S_n:=\,\vect[\F S_n]\,$.
	
	We define then $\:\displaystyle\C S:=\,\:\bigcup_{\mathclap{n\in \BB N}} \C S_n\:$, which we call the \emph{permutation algebra}.
}

Proposition \ref{open-represent} ensures that the $\,\C S_n\,$ and $\,\C S\,$ are $\ast$-algebras.

\section{Words and Observations}\label{sec_words}
	The representation of words over an alphabet in the unification algebra directly comes from the translation of Church lists in linear logic and their interpretation in Geometry of Interaction models~\cite{girard_geometry_1989,girard_geometry_1995}.

This proof-theoretic origin is an useful guide for intuition, even if we give here a more straightforward definition of the notion.

\medskip
From now on, we fix a set of two distinguished constant symbols $\,\IO:=\{\:\o,\i\:\}\,$.

\define[word algebra]
	{To a set $\,S\,$ of closed terms, we associate the $\ast$-algebra
	
	\vbox{
	\[S^\ast:= \vect\set{t\flow u}{t,u\in S}\]
	
	\hfill{\it\scriptsize(which is indeed an algebra because unification of closed terms is simply equality)}
	}
	
	\smallskip
	The \emph{word algebra} associated to a finite set of constant symbols $\,\Sigma\,$ is the $\ast$-algebra defined as 
	
	\vbox{
	\[\,\C W_\Sigma:=(\unitalg\ptensor\Sigma^\ast\ptensor\IO^\ast)\ptensor(\closed^\ast)\tpower 1\,\]

	\vspace{-1mm}
	\begin{flushright}
		{\it\scriptsize($\,\closed\,$ is the set of all closed terms, $\,\unitalg\,$ is defined at the beginning of Sect.\ref{permutation} \\ $\,\ptensor\,$ is as in Definition \ref{ptensor} and $(.)\tpower 1$ is the case $n=1$ of Definition \ref{unbounded})}

	\end{flushright}
		}
	\vspace{-2mm}
}

The words we consider are cyclic, with a begin/end marker $\,\start\,$, a reserved constant symbol.
For example the word $\,\TT{0010}\,$ is to be thought of as $\,\start\TT{0010}=\TT{10}\!\start\!\TT{00}=\TT0\!\start\!\TT{001}=\cdots\:$.

We consider therefore that the alphabet $\,\Sigma\,$ always contains the symbol $\,\start\,$.

\define[word representation]
	{\label{words}Let $\,W=\start\TT c_1\dots \TT c_n\,$ be a word over $\,\Sigma\,$ and $\,t_0,t_1,\dots,t_n\,$ be distinct closed terms.
	
	The \emph{representation} $\,W(t_0,t_1,\dots,t_n)\in \C W_\Sigma^\concrete\,$
	with respect to $\,t_0,t_1,\dots,t_n\,$
	of $\,W\,$ is an isometric wiring (Definition~\ref{concrete}), defined as
	\vspace{-1mm}	
\[
	\begin{array}{ccl}
		W(t_0,t_1,\dots,t_n) :=		& &x\p\start\p\i\p (t_0\p y) \sflow x\p\TT c_1\p\o\p (t_1\p y) \\
				&+& x\p\TT c_1\p\i\p (t_1\p y) \sflow x\p\TT c_2\p\o\p (t_2\p y) \\
				&+&\ \cdots\ \\
				&+& x\p\TT c_n\p\i\p (t_n\p y) \sflow x\p\start\p\o\p (t_0\p y)
	\end{array}
\]
\vspace{-5mm}
}

	We now define \emph{observations}, programs computing on representations of words.
They  lie in a particular $\ast$-algebra based on the representation of permutations presented in Sect.\ref{permutation}.

\define[observation algebra]
	{An \emph{observation} over a finite set of symbols $\,\Sigma\,$ is any element of $\,\C O_\Sigma^\concrete\,$ where $\,\C O_\Sigma:=\,(\closed^\ast\ptensor\Sigma^\ast\ptensor\IO^\ast)\ptensor\C S\,$, \textit{i.e.} a \emph{finite} sum of flows of the form
		\vspace{-1mm}
	\[(s'\p\TT c' \p \TT d' \flow s\p\TT c\p\TT d)\ptensor[\sigma]\]
	\vspace{-4mm}
	
with $\,s,s'\,$ closed terms, $\,\TT c,\TT c'\in\Sigma\,$, $\,\TT d,\TT d'\in \IO\,$ and $\,\sigma\,$ is a permutation.

\smallskip
Moreover when an observation happens to be an isometric wiring, we will call it an \emph{isometric observation}.
}

	\vspace{-2mm}

\section{Normativity: Independence from Representations}\label{sec_normativity}
	We are going to define how observations accept and reject words.
This needs to be discussed, because there is an issue with word representations: an observation is an element of $\,\ualg\,$ and can therefore only interact with \emph{representations} of a word, and there are many possible representation of the same word (in Definition \ref{words}, different choices of closed terms lead to different representations).
Therefore one has to ensure that acceptance or rejection is independent of the representation, so that the notion makes the intended sense.

The termination of computations will correspond to the algebraic notion of \emph{nilpotency}, which we recall here.

\define[nilpotency]
{A wiring $\,F\,$ is \emph{nilpotent} if $\,F^n=0\,$ for some $\,n\,$.}

\define[automorphism]
	{An \emph{automorphism} of a $\ast$-algebra $\,\C A\,$ is a linear application $\,\varphi\,:\: \C A\rightarrow\C A\,$ such that for all~$\,F,G\in\C A\,$: $\:\varphi(FG)=\varphi(F)\varphi(G)\,$, $\,\varphi(F^\dagger)=\varphi(F)^\dagger\,$ and $\,\varphi\,\text{ is injective}\,$.

}

\example{$\,\varphi(U_1\ptensor U_2):=U_2\ptensor U_1\,$ defines an automorphism of $\,\ualg \ptensor\ualg\,$.}

\smallskip
\notation{If $\,\varphi\,$ is an automorphism of $\,\C A\,$ and $\,\psi\,$ is an automorphism of $\,\C B\,$, we write $\,\varphi\ptensor\psi\,$ the automorphism of $\,\C A\ptensor\C B\,$ defined for all $\,A\in \C A, B\in \C B\,$ as $\,(\varphi\ptensor\psi)(A\ptensor B):=\varphi(A)\ptensor\psi(B)\,$.}

\define[normative pair]
	{\label{normpair}A pair $\,(\C A,\C B)\,$ of $\ast$-algebras is a \emph{normative pair} whenever any automorphism $\,\varphi\,$ of $\,\C A\,$ can be extended into an automorphism $\,\overline\varphi\,$ of the $\ast$-algebra $\,\C E\,$ generated by $\,\C A \cup\C B\,$ such that $\,\overline\varphi(F)=F\,$ for any $\,F\in \C B\,\subseteq\,\C E\,$.
}
The two following propositions set the basis for a notion of acceptance/rejection independent of the representation of a word.

\proposition[automorphic representations]
	{\label{automorphic}Any two representations $\,W(t_0,\dots,t_n),W(u_0,\dots,u_n)\,$ of a word $\,W\,$ over $\,\Sigma\,$ are automorphic: there exists an automorphism $\,\varphi\,$ of $\,(\closed^\ast)\tpower 1\,$ such that \[\,(\Id_{\,\ualg}\ptensor\varphi)\big(W(t_0,\dots,t_n)\big)=W(u_0,\dots,u_n)\,\]
}
\vspace{-8mm}
\proof{Consider a bijection $\,f\,:\,\closed \rightarrow\closed\,$ such that $\,f(t_i)=u_i\,$ for all $\,i\,$. Then set $\,\varphi(v\p x\flow w\p x):=f(v)\p x\flow f(w)\p x\,$, extended by linearity. 
}
\vspace{-1mm}
\smallskip
\proposition[nilpotency and normative pairs]
	{\label{normative}Let $\,(\C A,\C B)\,$ be a normative pair and $\,\varphi\,$ an automorphism of $\,\C A\,$. Let
	$\,F\in \ualg\ptensor\C A\,$, $\,G\in \ualg\ptensor\C B\,$ and let $\,\psi:=\Id_\ualg\ptensor \varphi\,$. Then $\,GF\,$ is nilpotent if and only if $\,G\,\psi(F)\,$ is nilpotent.%
}
\vspace{-2mm}
\proof{Let $\,\overline\varphi\,$ be the extension of $\,\varphi\,$ as in Definition \ref{normpair} and $\,\overline\psi:=\Id_{\,\ualg}\ptensor\overline\varphi\,$.

We have for all $\,n\neq 0\,$ that $\,(G\psi(F))^n=(\overline\psi(G)\overline\psi(F))^n=(\overline\psi(GF))^n=\overline\psi((GF)^n)\,$.

By injectivity of $\,\overline\psi\,$, $\,(G\psi(F))^n=0\,$ if and only if $\,(GF)^n=0\,$.
}

\corollary[independence]
{\label{indep}If $\,\big((\closed^\ast)\tpower 1,\C B\big)\,$ is a normative pair, $\,W\,$ a word over $\,\Sigma\,$ and $\,F\in \ualg\ptensor\C B\,$. The product of $\,F\,$ with the representation of the word, $\,FW(t_0,\dots,t_n)\,$, is nilpotent for one choice of $\,(t_0,\dots,t_n)\,$ if and only if it is nilpotent for all choices of $\,(t_0,\dots,t_n)\,$.}


The basic components of the word and observation algebras we introduced earlier can be shown to form a normative pair.

\theorem
{The pair $\,\big((\closed^\ast)\tpower1,\C S\big)\,$ is normative.}

\proof[sketch]{By simple computations, the set
\[\,\C A:=\vect\set{\sigma F}{\sigma \in \C S \mbox{ and }\, F\in (\closed^\ast)\tpower\infty}\,\]
 can be shown to be a $\ast$-algebra $\,\C E\,$, the $\ast$-algebra generated by $\,\C S\cup(\closed^\ast)\tpower 1\,$.

As $\,\varphi\,$ is an automorphism of $\,(\closed^\ast)\tpower 1\,$, it can be written as $\,\varphi(G\ptensor \unit)=\psi(G)\ptensor \unit\,$ for all $\,G\,$, with $\,\psi\,$ an automorphism of $\,\closed^\ast\,$.

We set for $\,F=F_1\ptensor\cdots\ptensor F_n\ptensor \unit\in(\closed^\ast)\tpower n\,$, $\,\tilde\varphi(F):=\psi(F_1)\ptensor\cdots\ptensor \psi(F_n)\ptensor \unit\,$ which extends into an automorphism of $\,(\closed^\ast)\tpower \infty\,$ by linearity.
Finally, we extend $\,\tilde\varphi\,$ to $\,\C A\,$ by $\,\overline\varphi(\sigma F):=\sigma\,\tilde\varphi(F)\,$.  It is then easy to check that $\,\overline\varphi\,$ has the required properties.
}

\medskip
\remark{Here we sketched a direct proof for brevity, but this can also be shown by involving a little more mathematical structure (actions of permutations on the unbounded tensor and crossed products) which would give a more synthetic proof.}

\medskip
We can then define the notion of the language recognized by an observation, \textit{via} Corollary \ref{indep}.

\define[language of an observation]
	{Let $\,\phi\in\C O_\Sigma^+\,$ be an observation over $\,\Sigma\,$.
	The \emph{language recognized by $\,\phi\,$} is the following set of words over $\,\Sigma\,$:
	\vspace{-1mm}
	\[\lang(\phi):=\set{W\:\mbox{word over}\,\Sigma}{\phi\,W(t_0,\dots,t_n)\,\mbox{nilpotent for any }\,(t_0,\dots,t_n)}\]
	}
	\vspace{-10mm}

\section{Wirings and Logarithmic Space}\label{sec_logspace}
	Now that we have defined our framework and showed how observations can compute, we study the complexity of deciding whenever an observation accepts a word (\ref{subsec_soundness}), and how wirings can decide any language in \textsc{(N)Logspace} (\ref{subsec_completness}).
	\subsection{Soundness of Observations}
	\label{subsec_soundness}
	The aim of this subsection is to prove the following theorem:

\vbox{
\theorem[space soundness]
	{\label{soundness}Let $\,\phi\in\C O_\Sigma^+\,$ be an observation over $\,\Sigma\,$.
	\begin{itemize}
		\item $\lang(\phi)\,$ is decidable in non-deterministic logarithmic space.
		\item If $\,\phi\,$ is isometric, then $\lang(\phi)\,$ is decidable in deterministic logarithmic space.
	\end{itemize}
}
}
Actually, the result stands for the complements of these languages, but as {\sc co-NLogspace = NLogspace} by the Immerman-Szelepcsényi theorem, this makes no difference.

\smallskip
The main tool for this purpose is the notion of \emph{computation space}: finite dimensional subspaces of $\,\closedh\,$ (Definition \ref{act_vect}) on which we will be able to observe the behavior of certain wirings. It can be understood as the place where all the relevant computation takes place.

\define[separating space]
{A subspace $\,E\,$ of $\,\closedh\,$ is \emph{separating} for a wiring $\,F\,$ whenever $\,F(E)\subseteq E\,$ and $\,F^n(E)=0\,$ implies $\,F^n=0\,$.}

Observations are \emph{finite} sums of wirings. We can naturally associate a finite-dimensional vector space to an observation and a finite set of closed terms.

\define[computation space]
{\label{compspace}Let $\,\{\,t_0,\dots, t_n\,\}\,$ be a set of distinct closed terms and $\,\phi\in\C O_\Sigma^\concrete\,$ an observation.

Let $\,N(\phi)\,$ be the smallest integer and $\,\TT S(\phi)\,$ the smallest (finite) set of closed terms such that $\,\phi \in (\TT S(\phi)^\ast\ptensor \Sigma^\ast\ptensor\IO^\ast)\ptensor\C S_{N(\phi)}\,$.

\smallskip
The \emph{computation space} $\,\comp_\phi (t_0,\dots, t_n)\,$ is the subspace of $\,\closedh\,$ generated by the terms
\[s\p \TT c\p \TT d \p (\,a_1\p\,\cdots\,\p a_{N(\phi)}\,\p\, \start)\]
where $\,s\in \TT S(\phi)\,$, $\,\TT c\in \Sigma\,$, $\,\TT d\in\IO\,$ and the $\,a_i\in \{\,t_0,\dots, t_n\,\}\,$.

The dimension of $\,\comp_\phi (t_0,\dots, t_n)\,$ is $\,|\Sigma| 2 (n+1)^{N(\phi)}|\TT S(\phi)|\,$ (where $\,|A|\,$ is the cardinal of $\,A\,$), which is polynomial in $\,n\,$. 
}

\lemma[separation]
{\label{sep}For any observation $\,\phi\,$ and any word $\,W\,$, the space $\,\comp_\phi(t_0,\dots, t_n)$ is separating for the wiring $\,\phi \,W(t_0,\dots, t_n)\,$.}

\proof[of Theorem~\ref{soundness}]{With these lemmas at hand, we can define the non-deterministic algorithm below. It takes as an input the representation $\,W(t_0,\dots, t_n)\,$ of a word $\,W\,$ of length $\,n\,$.

$\phi\,$ being a constant, one can compute once and for all $\,N(\phi)\,$ and $\,\TT S(\phi)\,$.

\begin{multicols}{2}
\begin{algorithmic}[1]
\STATE $D\gets 2|\TT S(\phi)|\,|\Sigma|(n+1)^{N(\phi)}$
\STATE $C\gets 0$
\STATE pick a term $\,v\in\comp_\phi(t_0,\dots,t_n)\,$\label{pick1}
\WHILE{$C\leq D$}
	\IF{$(\phi W(t_0,\dots,t_n))(v)=0$} \label{if}
		\RETURN ACCEPT
	\ENDIF
	\STATE pick a term $\,v'\,$ \\in $\,(\phi W(t_0,\dots,t_n))(v)\,$ \label{pick2}
	\STATE $v\gets v'$
	\STATE $C\gets C+1$
\ENDWHILE
\RETURN REJECT
\end{algorithmic}
\end{multicols}

\smallskip
All computation paths (the \enquote{pick} at lines \ref{pick1} and \ref{pick2} being non-deterministic choices) accept if and only if $\,(\phi W(t_0,\dots,t_n))^n(\comp_\phi(t_0,\dots,t_n))=0\,$ for some $\,n\,$ lesser or equal to the dimension $D$ of the computation space $\,\comp_\phi(t_0,\dots,t_n)\,$.
By Lemma \ref{sep}, this is equivalent to $\,\phi W(t_0,\dots,t_n)\,$ being nilpotent.

The term chosen at lines \ref{pick1} is representable by an integer of size at most $D$ and is erased by the one chosen at line \ref{pick2} every time we go through the {\bf while}-loop.
$C$ and $D$ are integers proportional to the dimension of the computation space, which is representable in logarithmic space in the size of the input (Definition~\ref{compspace}).

\smallskip
The computation of $\,(\phi W(t_0,\dots,t_n))(v)\,$ at line \ref{if} and \ref{pick2} and can be performed in logarithmic space by Proposition \ref{unif-logspace}, as we are unifying closed terms with linear terms.

\smallskip
Moreover, if $\,\phi\,$ is an isometric wiring, $\,(\phi W(t_0,\dots,t_n))(v)\,$ consists of a single term instead of a sum by Lemma \ref{lem_isom}, and there is therefore no non-deterministic choice to be made at line \ref{pick2}.
It is then enough to run the algorithm enumerating all possible terms of $\,\comp_\phi(t_0,\dots,t_n)\,$ at line \ref{pick1} to determine the nilpotency of $\,\phi W(t_0,\dots,t_n)\,$.
}

	\subsection{Completeness: Representing Pointer Machines as Wirings}
	\label{subsec_completness}
	To prove the converse of Theorem \ref{soundness}, we prove that wirings can encode a special kind of read-only multi-head Turing Machine: pointers machines.
The definition of this model will be guided by our understanding of the computation of wirings: they won't have the ability to write and acceptance will be defined as termination of all paths of computation.
For a survey of this topic, one may consult the first author's thesis \cite[Chap.4]{Aubert2013b}, the main novelty of this part of our work is to notice that reversible computation is represented by isometric operators.



\define[pointer machine]
{A \emph{pointer machine} over an alphabet $\,\Sigma\,$ is a tuple $\,(N,\TT S,\Delta)\,$ where
\begin{itemize}
	\item $N\neq 0\,$ is an integer, the \emph{number of pointers},
	\item $\TT S\,$ is a finite set, the \emph{states} of the machine,
	\item $\Delta \,\subseteq\, (\TT S\times\Sigma\times\IO)\times(\TT S\times\Sigma\times\IO)\times \F S_N\:$, the \emph{transitions} of the machine
	
	(we will write $\,(s,\TT c,\TT d) \rightarrow (s',\TT c',\TT d') \times \sigma\,$ the transitions, for readability).
\end{itemize}
A pointer machine will be called \emph{deterministic} if for any $\,A \,\in\, \TT S\times \Sigma\times \IO\,$, there is at most one $\,B\,\in\, \TT S\times \Sigma\times \IO\,$ and one $\,\sigma\in \F S_N\,$ such that $\,A\rightarrow B \times \sigma\,\in\,\Delta\,$.
In that case we can see $\,\Delta\,$ as a partial function, and we say that $\,M\,$ is \emph{reversible} if $\,\Delta\,$ is a partial injection.
}
We call the first of the $\,N\,$ pointers the \emph{main} pointer, it is the only one that can move.
The other pointers are referred to as the \emph{auxiliary} pointers.
An auxiliary pointer will be able to become the main pointer during the computation thanks to permutations.

\define[configuration]
{Given the length $\,n\,$ of a word $\,W=\start\TT c_1\dots \TT c_n\,$ over $\,\Sigma\,$ and a pointer machine $\,M=(N,\TT S,\Delta)\,$, a \emph{configuration} of $\,(M,n)\,$ is an element of 
\[\,\TT S\times\Sigma\times\IO\times\{0,1,\dots,n\}^N\,\]}
\vspace{-5mm}
The element of $\,\TT S\,$ is the state of the machine and the element of $\,\Sigma\,$ is the letter the main pointer points at.
The element of $\,\IO\,$ is the direction of the next move of the main pointer, and the elements of $\,\{0,1,\dots,n\}^N\,$ correspond to the positions of the (main and auxiliary)  pointers on the input.

\smallskip

As the input tape is considered cyclic with a special symbol marking the beginning of the word (recall Definition \ref{words}), the pointer positions are integers \emph{modulo} $\,n+1\,$ for an input word of length $\,n\,$.

\define[transition]
{Let $\,W\,$ be a word and $\,M=(N,\TT S,\Delta)\,$ be a pointer machine.
A \emph{transition} of $\,M\,$ on input $\,W\,$ is a triple of configurations
\[s,\TT c,\TT d,(p_1,\dots,p_N) \trans{\TT{MOVE}}{} s,\TT c',\overline{\TT d},(p_1',\dots,p_N') \trans{\TT{SWAP}}{} s',\TT c'',\TT d',(p_{\sigma(1)}',\dots,p_{\sigma(N)}') \]
such that
\begin{enumerate}
	\item if $\,\TT d\in\IO\,$, $\,\overline{\TT d}\,$ is the other element of $\,\IO\,$,
	\item $p_1'=p_1+1\,$ if $\,\TT d=\i\,$ and $\,p_1'=p_1-1\,$ if $\,\TT d=\o\,$,
	\item $p_i'=p_i\,$ for $\,i\neq 1\,$,
	\item $\TT c\,$ is the letter at position $\,p_1\,$ and $\,\TT c'\,$ is the letter at position $\,p_1'\,$, \label{condition}
	\item and $(s,\TT c',\overline{\TT d}) \rightarrow (s',\TT c'',\TT d') \times \sigma\,$ belongs to $\,\Delta\,$.
\end{enumerate}
}
There is no constraint on $\,c''\,$, but every time this value differs from the letter pointed by $\,p_{\sigma(1)}'\,$, the computation will halt on the next \texttt{MOVE} phase, because there is a mismatch between the value that is supposed to have been read and the actual bit of $\,W\,$ stored at this position, and that would contradict the first part of item \ref{condition}.
In terms of wirings, the \texttt{MOVE} phase corresponds to the application of the representation of the word, whereas the \texttt{SWAP} phase corresponds to the application of the observation.


\define[acceptance]
{\label{translate}We say that $\,M\,$ accepts $\,W\,$ if any sequence of transitions $\,\big(C_i\trans{\TT{MOVE}}{}C_i'\trans{\TT{SWAP}}{}C_i''\big)\,$ such that $\,C''_i=C_{i+1}\,$ for all $\,i\,$ is necessarily finite.

We write $\,\lang(M)\,$ the set of words accepted by $\,M\,$.}

This means informally: we consider that a pointer machine accepts a word if it cannot ever loop, from whatever configuration it starts from.
That a lot of paths of computation accepts \enquote{wrongly} is no worry, since only rejection is meaningful: our pointer machines compute in a \enquote{universally non-deterministic} way, to stick to the acceptance condition of wirings, nilpotency.



\proposition[space and pointer machines]
{\label{pointerl}If $\,L\in \text{\sc NLogspace}\,$, then there exist a pointer machine $\,M\,$ such that $\,\lang(M)=L\,$.
Moreover, if $\,L\in \text{\sc Logspace}\,$ then $\,M\,$ can be chosen to be reversible.
}
\proof[sketch]{%
It is well-known that read-only Turing Machines --~or equivalently (non-)Deterministic Multi-Head Finite Automata~-- characterize {\sc (N)Logspace} \cite{Hartmanis1972}.
It takes little effort to see that our pointer machines are just a reasonable rearrangement of this model, since it is always possible to encode the missing information in the states of the machine.

That acceptance and rejections are \enquote{reversed} is harmless in the deterministic (or equivalently reversible \cite{Lange2000}) case, and uses that {\sc co-NLogspace = NLogspace} to get the expected result in the non-deterministic case.
}

\bigskip
As we said, our pointer machines are designed to be easily simulated by wirings, so that we get the expected result almost for free.

\theorem[space completeness]
{If $\,L\in \text{\sc NLogspace}\,$, then there exist an observation $\,\phi\in\C O_\Sigma^+\,$ such that $\,\lang(\phi)=L\,$.
Moreover, if $\,L\in \text{\sc Logspace}\,$ then $\,\phi\,$ is an isometric wiring.
}

\proof{%
By Proposition \ref{pointerl}, there exists a pointer machine $\,M=(N,\TT S,\Delta)\,$ such that $\lang(M) = L$.
We associate to the set $\,\TT S\,$ a set of distinct closed terms $\,[\TT S]\,$ and write $\,[s]\,$ the term associated to $\,s\,$.
To any element $\,D=(s,\TT c,\TT d) \rightarrow (s',\TT c',\TT d') \times \sigma\,$ of $\,\Delta\,$ we associate the flow 
\[[D]:=([s']\p\TT c'\p\TT d' \flow [s]\p\TT c\p\TT d) \ptensor [\sigma] \:\in([\TT S]^\ast\ptensor\Sigma^\ast\ptensor\IO^\ast)\ptensor\C S_n\:\subseteq\,\C O_\Sigma^+\,\]
and we define the observation $\,[M]\in\C O_\Sigma^+\,$ as $\,\displaystyle\sum_{D\in \Delta} [D]\,$.

One can easily check that this translation preserves the language recognized (there is even a step by step simulation of the computation on the word $\,W\,$ by the wiring $\,[M]W(t_0,\dots,t_n)\,$) and relates reversibility with isometricity: in fact, $\,M\,$ is reversible if and only if $\,[M]\,$ is an isometric wiring.
Then, if $\,L\in \text{\sc Logspace}\,$, $\,M\,$ is deterministic and can always be chosen to be reversible \cite{Lange2000}.
}

\section*{Discussion}
 \addcontentsline{toc}{section}{Discussion}
The language of the unification algebra gives us a twofold point of view on computation, either through algebraic structures (that are described finitely by wirings) or pointer machines.
We may therefore start exploring possible variations of the construction, combining intuitions from both worlds.

\smallskip
For instance, the choice of a normative pair can affect the expressivity of the construction:
the more restrictive the notion of representation of a word is, the more liberal that of an observation can become, as suggested by T.~Seiller.
Whether and how this can affect the corresponding complexity class is definitely a direction for future work.



\smallskip
Another pending question about this approach to complexity classes is to delimit the minimal prerequisites of the construction, its core.

Earlier works \cite{girard_normativity_2012,aubert_characterizing_2012,seiller_logarithmic_2013} made use of von Neumann algebras to get a setting that is expressive enough, we ligthen the construction by using simpler objects.
Yet, the possibility of representing the action of permutations on a unbounded tensor product is a common denominator that seems deeply related to logarithmic space and pointer machines.

\smallskip
The logical counterpart of this work also needs clarifying.
Indeed, the idea of representation of words comes directly from proof-theory, while the notion of observation does not seem to correspond to any known logical construction.

\smallskip
Finally, execution in our setting being based on iteration of matching, which is computable efficiently by a parallel machine, it seems possible to relate our modelisation with parallel computation.

\newpage

\bibliographystyle{splncs}
\bibliography{girard,complexity,phd,unification,goi}
\addcontentsline{toc}{section}{References}

\end{document}